\documentclass[12pt]{article}
\usepackage{graphicx}

\begin{document}

%
\catcode`\@=11
\@addtoreset{equation}{section}
\def\theequation{\thesection.\arabic{equation}}
\catcode`\@=12
\newcommand{\be}{ \begin{equation}}
\newcommand{\ee}{ \end{equation}  }
\newcommand{\bea}{ \begin{eqnarray}}
\newcommand{\eea}{ \end{eqnarray}  }
\newcommand{\bi}{\bibitem}
\newcommand{\rd}{ \mbox{\rm d} }
\newcommand{\rD}{ \mbox{\rm D} }
\newcommand{\re}{ \mbox{\rm e} }
\newcommand{\rO}{ \mbox{\rm O} }
\newcommand{\erf}{\mbox{\rm erf}}
\newcommand{\diag}{\mbox{\rm diag}}

\renewcommand{\floatpagefraction}{0.8}

\def\del{\partial}
\def\phat{\hat{p}}
\def\Delbar{\bar{\Delta}}
\def\SF{Schr\"odinger functional }

\def\rmd{{\rm d}}
\def\rmD{{\rm D}}
\def\rme{{\rm e}}
\def\rmO{{\rm O}}
\def\tr{{\rm tr}}

 
\def\gms{g_{\ms}}
\def\gmsbar{g_{\msbar}}
\def\gbar{\bar{g}}
\def\gup{g_{\Upsilon}}
\def\gbarms{\gbar_{\ms}}
\def\gbarmom{\gbar_{\rm mom}}
\def\ms{{\rm MS}}
\def\msbar{{\rm \overline{MS\kern-0.14em}\kern0.14em}}
\def\lat{{\rm lat}}
\def\glat{g_{\lat}}
\def\gSF{g_{\hbox{\rm \scriptsize SF}}}

\def\alphabar{\alpha}
\def\alphasf{\alpha_{\rm SF}}
\def\alphat{\tilde{\alpha}_0}

\def\tm{{\cal T}}


\begin{titlepage}

%

\vspace{1.cm}
\begin{center}
{\LARGE Helicity modulus as renormalized coupling\\ in the O(3) $\sigma$-model}
\vskip 1 cm
\vbox{
\centerline{
\includegraphics[width=2.5cm]{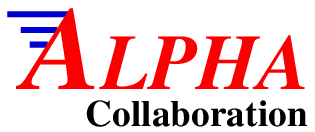}
}
}
\vskip 1 cm

{\large Heiko Molke\footnote{
Present address: DESY, Platanenallee 6, D-15738 Zeuthen
}
 and Ulli Wolff\\
Institut f\"ur Physik, Humboldt Universit\"at\\
Invalidenstr. 110, D-10099 Berlin, Germany\\}
\end{center}
\vspace{.5cm}
\thispagestyle{empty}
\begin{abstract}\normalsize
For the  family of O($n$) invariant nonlinear $\sigma$-models
we consider boundary conditions that are periodic up to an  
O($n$) rotation. The helicity modulus is related
to the change in free energy under variations
of the corresponding angle. It defines a nonperturbative finite
volume running coupling similar to the Schr\"odinger functional
for QCD. For the two-dimensional O(3)-model we investigate  this quantity by
analytical and numerical techniques. We establish its universal 
continuum relation
to the finite volume massgap coupling at all scales and coupling strengths.

\end{abstract}

\end{titlepage}

\section{Introduction}

The property of asymptotic freedom is a decisive
feature of QCD as well as of a large class of
two dimensional nonabelian spin models like the O($n$)
$\sigma$-models for $n>2$. 
Although it is based only on proofs in
perturbation theory (to all orders), the following structural properties of
these models are widely accepted and assumed here\footnote{
See ref.~\cite{HERESY} for a diverging point of view
}.
The continuum limit is reached at vanishing bare coupling
and a mass scale emerges  in the renormalized continuum theory
by dimensional transmutation.
Many features associated with much higher energies or short distance
can be related to each other by renormalized
perturbation theory. Usually one singles out one suitable
high energy quantity as renormalized coupling and 
constructs expansions for other observables in its powers.
Other phenomena around the fundamental scale, like the spectrum,
are not accessible to
this strategy and are often investigated by numerical simulation.
These opposite ``sectors'' are really features of one and the same
theory and it is hence both interesting and possible to relate them,
that is, to compute renormalized coupling constants
at high energy in terms of the low energy scale.
This has been the programme of the ALPHA collaboration in recent years.
An efficient method has been developed first for the O(3) $\sigma$-model
\cite{LWW} followed by quenched QCD which is reviewed in \cite{REV}.
Many present activities are related to the goal of including dynamical quarks.

To relate the perturbative sector with low energy
physics very dissimilar continuum scales have to be accommodated
on a lattice with a spacing that is small compared to all other
scales. In a direct approach this either calls for unmanageably large
lattices or one has to compromise with the attempted limits like
the continuum extrapolation. The ALPHA strategy overcomes this
problem by a finite size scaling technique. A running coupling
constant $\gbar^2(L)$ is constructed in the continuum limit,
which at large $L$ can be related to spectral scales and
which at small $L$ can be used as an expansion parameter and thus
gives access to the perturbative sector. Step by step one computes 
$\gbar^2(2L)$ in terms of $\gbar^2(L)$ by continuum extrapolation.
Since the system size $L$ is used as the physical scale,
$L/a$ is the only large scale ratio, where $a$ is the lattice spacing.
The choice of $\gbar$ is not unique and a number of practical criteria
were taken into account. 

For the O($n$) model the finite volume
mass gap was used in \cite{LWW}
\be
\gbar^2 = \frac{2}{n-1} \, m(L) L ,
\label{LWWcoupling}
\ee
where $m(L)$ is the gap of the transfer matrix referring to
spatial periodic boundary conditions with periodicity $L$. 
For $L\to\infty$,  $m(L)$ saturates to the infinite volume massgap,
which we identify with the dynamically generated scale.
At small $L$ the mass gap becomes perturbative \cite{MLML} and can be used 
as an
expansion parameter.
For gauge theories a similarly convenient though less obvious quantity
was defined via the \SF \cite{LNWW, REV}. The basic mechanism is
to study the response of the free energy under the variation
of an angle that enters into non-trivial boundary conditions.
Again the system size is the only scale beside the cutoff,
and for the quenched theory $\alpha_s$ in the high energy limit 
has been computed
to a satisfactory precision.

In this article we define and investigate an alternative coupling
$\gup(L)$ for the O($n$) $\sigma$-model in two dimensions,
which is closely related to the helicity modulus.
We compare its properties with the massgap coupling $\gbar(L)$.
They can be analytically related in the continuum for both small
and large coupling. Both expansions are checked 
and the crossover
range is controlled by high precision numerical simulations.
In the next section we define the helicity modulus
and relate it to $\gbar$ at strong coupling.
In section 3 $\gup$ is properly normalized and its weak coupling
behavior is explored up to two loops with details given in appendix~A.
Section 4 summarizes our numerical work and gives conclusions.
This work is based on the diploma thesis \cite{HM} of H.~Molke.
The introduction of $\gup$ goes back to earlier attempts \cite{UW1,UW2}
to investigate renormalization by finite size techniques.

\section{Helicity modulus and transfer matrix}

In this section we introduce the helicity modulus.
For earlier discussions of similar quantities see Ref.~\cite{helmod}
and further references found there.

We consider the O($n$) $\sigma$-model with its standard nearest neighbor
lattice action
\be
S = - \frac{1}{g_0^2} \sum_{x \mu} s(x) s(x+a\hat{\mu}),
\ee
where $s(x)$ is the unit length spin field and $\hat{\mu}, \mu=0,1$, are
unit vectors along the axes of a square lattice
with spacing $a$. We take $T$ as the size in the time or 0-direction and
$L$ in the other direction. In space direction we impose strictly
periodic boundary conditions, while in the time direction we demand
periodicity up to a planar SO($n$) rotation in spin space by an angle $\alpha$,
\be
s(x+T\hat{0}) = \exp(\alpha K_{ij}) s(x).
\ee
Here $K_{ij}$ generates rotations in the $ij$ plane,
\be
\Bigl( K_{ij} \Bigr)_{kl} = \delta_{ik} \delta_{jl} - \delta_{il}
\delta_{jk}\, ,
\label{Ongen}
\ee
and we assume $i \not= j$ fixed to an arbitrary pair of values until
further notice.
Integration of all spins 
with the O($n$) invariant measure gives the partition function
\be
Z_{\alpha} = \int Ds \exp(-S).
\label{funcint}
\ee
Ratios $Z_{\alpha_1}/Z_{\alpha_2}$ depend on differences in free energy
for different boundary conditions and are expected to be universal.
We now define the helicity modulus $\Upsilon$ by
\be
\Upsilon =  - \frac1{Z_0} \left.
\frac{\del^2 Z_{\alpha}}{\del\alpha^2}
\right|_{\alpha=0}.
\ee

From this definition it is rather easy to establish a connection
between $\Upsilon$ and the transfer matrix $\tm$.
For the fully periodic case $Z_0$ we have
\be
Z_0 = \tr[\tm^{T/a}].
\ee
The extra twist by an angle $\alpha$ corresponds to inserting 
a rotation operator under the trace in the Hilbert space where
$\tm$ operates,
\be
 Z_\alpha = \tr\, [\tm^{T/a} \exp(i\alpha {\cal K}_{ij})].
\label{helmodtm}
\ee
If we realize states as wavefunctions $\psi(\sigma)$ on spatial 
one dimensional spin fields $\sigma$, this induced operator is given by
\be
\Bigl(\exp(i\alpha {\cal K}_{ij}) \psi \Bigr)(\sigma) = 
\psi\Bigl(\exp(-\alpha K_{ij}) \sigma\Bigr).
\ee
The operator $\tm$ possesses real positive eigenvalues
$\lambda_0 > \lambda_1 \ge \lambda_2 \ge \ldots $ with $\lambda_0$
corresponding to the nondegenerate ground state. These eigenvalues
depend on $L$ and $a$.
We define the finite volume massgap $m(L)$ as
\be
\exp(-m(L)a) = \frac{\lambda_1}{\lambda_0}.
\ee
Due to the O($n$) invariance $\tm$ and ${\cal K}_{ij}$ commute
and we simultaneously diagonalize the generator that appears in 
(\ref{helmodtm}).
Hence for each eigenstate there is a value $\mu_k, k=0,1,2,\ldots$. 
For $n=3$ these
are just the ``magnetic'' quantum numbers of eigenstates in all
possible integer isospin multiplets in the spectrum.
The $\alpha$-dependent partition function is now given by
\be
 Z_\alpha =  \sum_{k\ge0} (\lambda_k)^{T/a} \exp(i\alpha \mu_k),
\ee
and for the helicity modulus we obtain
\be
\Upsilon = \frac1{Z_0} \sum_{k\ge0} (\lambda_k)^{T/a}  \mu_k^2 \, .
\label{Upstates}
\ee
For large $mT$ we expect the ground state and the lowest excitations above
it to dominate. While the ground state is O($n$) invariant, $\mu_0=0$,
we expect an $n$-fold degenerate vector multiplet of one particle states
as the 
next excitations above it. On it the generators are represented in the form
(\ref{Ongen}) and have eigenvalues $1,-1$ and $n-2$ times 0.
Asymptotically this implies
\be
\Upsilon \stackrel{T\to\infty}{\simeq} 2 \sum_p \exp(-E_p T).
\ee
The sum here is over single particle momenta $p=2\pi j/L$ with 
$-L/2a < j \le L/2a$, and $E_p$ are the associated energies.
Contributions of higher states are assumed to be exponentially
suppressed. We now take the continuum limit
at fixed large values for $m L$ and $m T$.
The one-particle spectrum is expected to become relativistic,
$E_p=\sqrt{m^2+p^2}$ and we find that
$\Upsilon$ is given in terms
of the renormalized coupling (\ref{LWWcoupling}) introduced in \cite{LWW} by
\be
\Upsilon \stackrel{\gbar^2\gg 1}{\simeq} 
2 \sum_{j=-\infty}^{j=+\infty} 
\exp\left(-\rho\sqrt{\gamma^2+(2\pi j)^2 } \, \right).
\label{Upsc}
\ee
with
\be
\gamma = \frac{n-1}{2}\gbar^2.
\ee
This formula holds in the continuum limit for large $\gbar^2$ and fixed 
aspect ratio 
\be
\rho=\frac{T}{L}\, .
\ee

For not extremely large $\gbar^2$
it only takes the few lowest $j$-values to carry out the sum
in (\ref{Upsc}) numerically to machine precision. By Poisson resummation
one may derive the asymptotic form
\bea
\Upsilon &\simeq& \frac{2\gamma}{\pi} \, K_1(\rho\gamma) +
\rO\left(\exp(-\sqrt{\rho^2+1} \, \gamma) \right)\\
&\simeq& 
\sqrt{\frac{2\gamma}{\pi\rho}} \, \exp(-\rho\gamma)
(1+\rO(\gamma^{-1})),
\eea
where $K_1$ is a modified Bessel function.
In Fig.~1 we show how these asymptotic forms approach the sum
(\ref{Upsc}) for the case $\rho=1$. Also shown is the $j=0$ term alone,
which underestimates the sum by at most 5~\% for $\gamma \le 3.5$.
\begin{figure}
  \begin{center}
    \includegraphics[width=12cm]{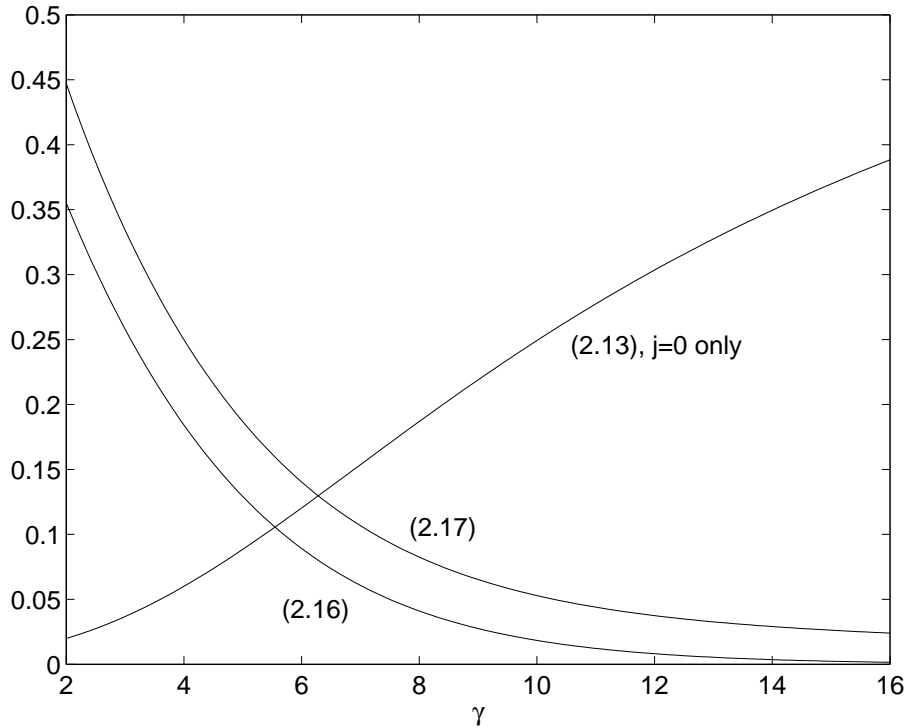}
  \end{center}
\caption{Relative discrepancies between the strong 
coupling formula (\ref{Upsc}) and approximations to it.}
\end{figure}

\section{Helicity modulus in perturbation theory}

\subsection{Preparation and leading order}

By changing the integration variables in (\ref{funcint})
\be
s(x) \to \exp\left(-\alpha\frac{x_0}{T}K_{ij} \right) s(x)
\ee
we arrive at
\be
Z_{\alpha} = \int Ds \exp(-S_B)
\ee
where the action
\be
S_B = - \frac{1}{g_0^2} \sum_{x \mu} s(x) B_{\mu} s(x+a\hat{\mu})
\ee
has acquired a constant SO($n$) gauge field
\be
B_0 = \exp\left(\alpha \frac{a}{T} K_{ij}\right), \quad B_1=1, 
\ee
and $s(x)$ has become strictly periodic.
The helicity modulus is now expressed by an expectation value
in the periodic ensemble
\be
\Upsilon = \left\langle\; \frac{\del^2 S_B}{\del\alpha^2} - 
\left( \frac{\del S_B}{\del\alpha} \right)^2  \;\right\rangle
\ee
with $B_{\mu}=1$ after taking all derivatives.
So far the angle $\alpha$ has referred to rotations with one
particular generator $K_{ij}$. At this point we average over
all planes $i<j$ and split $\Upsilon$ into two O($n$) invariant
contributions $\Upsilon=\Upsilon_1-\Upsilon_2$,
\be
\Upsilon_1 = \left\langle\; 
\frac{\del^2 S_B}{\del\alpha^2} \;\right\rangle =
\frac{2}{n\rho g_0^2}\; \frac{a^2}{V} 
\sum_x\, \langle s(x) s(x+a\hat{0}) \rangle
\label{U1def}
\ee
and
\be
\Upsilon_2 = \left\langle\; 
\left( \frac{\del S_B}{\del\alpha} \right)^2 \;\right\rangle =
\frac{2}{n(n-1)\rho g_0^4}\; \frac{a^4}{V} 
\sum_{i<j}\sum_{x y}\, \langle j^0_{ij}(x)j^0_{ij}(y)  \rangle.
\label{U2def}
\ee
In these expressions the volume $V=TL$ and the currents
\be
j^{\mu}_{ij}(x) = s(x) K_{ij}\Delta_\mu s(x)
\ee
with the discrete derivative
\be
\Delta_\mu s(x) = \frac{1}{a} \left[s(x+a\hat{\mu}) -s(x)\right]
\ee
have been introduced. 
While $\Upsilon_1$ is a nearest neighbor correlation
essentially corresponding to the internal energy, $\Upsilon_2$
is a kind of susceptibility with correlations over all separations. 
In perturbation theory contributions to
$\Upsilon_2$ start at the one loop level and we
find to leading order in $g_0$
\be
\Upsilon = \frac{2}{n\rho g_0^2} + \rO(1).
\ee
A correctly normalized and nonperturbatively defined renormalized
$L$-dependent coupling constant can now be defined as
\be
\gup^2 = \frac{2}{n\rho} \frac{1}{\Upsilon},
\label{gupdef}
\ee
quite in the spirit of the Schr\"odinger functional.
Its relation to $\gbar$ from ref.\cite{LWW} is 
\be
\gup^2 = \left\{\begin{tabular}{l@{ for }l}
$\gbar^2+\rO(\gbar^4)$ & $\gbar^2\to 0$ \\[1ex]
$\frac{1}{n \gbar} \sqrt{\frac{4\pi}{\rho(n-1)}}
\exp\left(\frac{n-1}{2}\rho\gbar^2\right) $
                &  $\gbar^2\to \infty$.
\end{tabular}\right.
\label{asymptotics}
\ee
As $\gbar^2$ is proportional to $L$ at large volume or strong coupling,
we find exponential growth for $\gup^2$. This is also expected
for the Schr\"odinger functional coupling in gauge theory \cite{SFSU2}.
The origin in both cases is the exponentially small sensitivity
of the free energy to boundary effects in a physically large volume.

\subsection{Results of a two loop calculation}

For simplicity we confined ourselves to the case $T=L, \rho=1$
for our perturbative and our numerical calculations.
Details on the perturbative evaluation of $\Upsilon$ are given in
appendix~A. Nontrivial coefficients were evaluated for sequences of
lattices of finite $L/a$ and then fitted to the expected asymptotic
form. The extrapolation was carried out as
described in appendix~D of ref.~\cite{BWW} with lattices up to
$L/a=100$. The cost in CPU time was negligible in this two dimensional case. 
As for the 
Schr\"odinger functional, it was advantageous to compute
some of the two loop diagrams in position space rather than momentum
space.

The nearest neighbor correlation in $\Upsilon_1$ is rather simple
to compute \cite{Epert} to the required order. 
As a result we find
\be
\Upsilon_1 = \frac{2}{n} \frac{1}{g_0^2} - \frac{n-1}{2n}
-\frac{n-1}{16n}\, g_0^2 + \rO(g_0^4,a^2).
\ee
The current-current correlation in $\Upsilon_2$ is more involved
and has the structure
\be
\Upsilon_2 = \Upsilon_2^{(0)} + \Upsilon_2^{(1)} g_0^2 + \rO(g_0^4).
\label{U2expand}
\ee

By combining the results for $\Upsilon_2^{(0)}$ and $\Upsilon_2^{(1)}$
from the appendix with the expansion of $\gbar^2$ on the lattice
\cite{LWW,DSS} we find in the continuum limit
the renormalized perturbative series
\be
\gup^2(L) = \gbar^2(sL) + d_1 \gbar^4(sL) + d_2 \gbar^6(sL) + \rO(\gbar^8(sL))
\label{gugb}
\ee
with
\bea
d_1(s) &=& (n-2)\, 0.16350689821 - \frac{1}{2\pi}\ln s
\label{d1s} \\
d_2(s) &=& d_1(s)^2 + (n-2)\,[\,0.00315826256 - (n-2)\, 0.007733893180\,]
\nonumber\\
&& - \frac{1}{(2\pi)^2}\ln s.
\label{dcoeff}
\eea
Note that  a free relative factor $s$ between the scales 
at which $\gup$ and $\gbar$ are taken has been introduced here for
later use.

\section{Numerical results}
Our main goal in this section is to establish the
nonperturbative relation between $\gbar^2(L)$ and $\gup^2(L)$
in the continuum limit of the $O(3)$ model
for arbitrary values of these couplings.
Our strategy is to first construct series of values $L/a$ and
$\beta=1/g_0^2$ which correspond to fixed values of $\gbar$.
For precisely the same series we measure $\gup^2$ and extrapolate
these values to $L/a\to\infty$. For the first part of the task
we include but extend as necessary the data from \cite{LWW}.
These runs were carried out with precisely the same code as described
there. 
In particular, we took advantage of free boundary conditions 
in the time direction to extract the massgap,
and the reweighting technique allowed for a post-run fine-tuning of $\beta$
to match a desired value of $\gbar^2$. The simulation of $\Upsilon$
is rather conventional on an $L\times L$ torus. We employed the single
cluster algorithm \cite{UW1C} and measured the observables given
in eqs.~(\ref{U1def}) and (\ref{U2def}).
Both kinds of results are collected in Table~\ref{alldata}.
\begin{table}[!ht]
 \begin{center}
 \begin{tabular}{|c|ccc|ccc|}\hline
 $L$      & $\beta$ & $\gbar^2$ & $\gup^2$ & $\beta$ &$\gbar^2$ & $\gup^2$  \\ \hline
  6       &         &           &          &1.8439 & 0.8166(5) & 0.9108(2) \\
  8       &         &           &          &1.8947 & 0.8166(4) & 0.9210(2) \\
 10       & 2.0489  & 0.7390(5) & 0.8276(1)&1.9319 & 0.8166(6) & 0.9276(2) \\
 12       &         &           &          &1.9637 & 0.8166(6) & 0.9295(2) \\ 
 16       & 2.1260  & 0.7390(6) & 0.8341(1)&2.0100 & 0.8166(5) & 0.9350(2) \\ 
 20       & 2.1626  & 0.7390(5) & 0.8361(1)&2.0489 & 0.8166(8) & 0.9351(2) \\
 24       & 2.1930  & 0.7390(6) & 0.8369(1)&       &          &          \\
 32       & 2.2422  & 0.7390(5) & 0.8363(1)&2.1260 & 0.8166(8) & 0.9375(3)  \\ \hline
  6       & 1.7276  & 0.9176(6) & 1.0327(3)&1.6050 & 1.0595(7) &1.2063(4) \\
  8       & 1.7791  & 0.9176(5) & 1.0460(3)&1.6589 & 1.0595(7) &1.2276(4) \\
 10       & 1.8171  & 0.9176(6) & 1.0551(3)&1.6982 & 1.0595(7) &1.2414(4) \\
 12       & 1.8497  & 0.9176(7) & 1.0587(3)&1.7306 & 1.0595(7) &1.2462(4) \\
 16       & 1.8965  & 0.9176(5) & 1.0642(3)&1.7800 & 1.0595(6) &1.2541(4) \\
 20       &         &           &          &1.8171 & 1.0595(7) &1.2579(4) \\ 
 24       & 1.9637  & 0.9176(6) & 1.0674(3)&       &           &          \\  \hline
 10       & 1.3634  & 2.0373(18)& 2.9635(12)&1.2939& 2.4596(16)&4.268(2) \\
 12       & 1.4060  & 2.0373(18)& 2.9848(12)&1.3413& 2.4596(16)&4.284(3) \\
 16       & 1.4683  & 2.0373(18)& 3.0120(12)&1.4095& 2.4596(17)&4.302(3) \\
 20       &         &           &           &1.4579& 2.4596(17)&4.316(3) \\
 24       & 1.5470  & 2.0373(18)& 3.0283(13)&1.4948& 2.4596(17)&4.337(3) \\
 32       & 1.6000  & 2.0373(18)& 3.0314(18)&1.5500& 2.4596(16)&4.344(3) \\ 
 \hline
 10       & 1.2211  & 3.0280(27)& 7.022(6)  &1.1427& 3.7692(17)&13.947(8)\\
 12       & 1.2736  & 3.0280(27)& 7.076(7)  &1.2014& 3.7692(19)&14.030(8)\\
 16       & 1.3481  & 3.0280(27)& 7.140(7)  &1.2847& 3.7692(18)&14.106(9)\\
 24       & 1.4413  & 3.0280(27)& 7.158(8)  &1.3862& 3.7692(17)&14.179(11)\\
 32       & 1.5000  & 3.0280(27)& 7.208(8)  &1.4500& 3.7692(16)&14.218(11)\\
 \hline
 \end{tabular}
 \caption{Simulation results for $\gbar^2$ and $\gup^2$.}
 \label{alldata}
 \end{center}
\end{table}

For each of the eight series at fixed $\gbar^2$ we have pairs of
values with errors $\delta$ of 
$\gbar^2$ and $(\delta_\Upsilon)$ of $\gup^2$.
For the extrapolation in $a/L$ we combine them into effective errors
in $\gup^2$ only 
\be
\Delta(\gup^2) = \sqrt{\delta_\Upsilon^2 + 
\left( \frac{\del\gup^2}{\del\gbar^2} \delta \right)^2} \; .
\ee
The required slope can be estimated with sufficient accuracy
from the weak and strong coupling behavior.
We then extrapolate by fitting a function $A+B (a/L)^2$ to the
$\gup^2$ values with these errors. 
\begin{figure}\label{extrapol}
  \begin{center}
    \includegraphics[width=12cm]{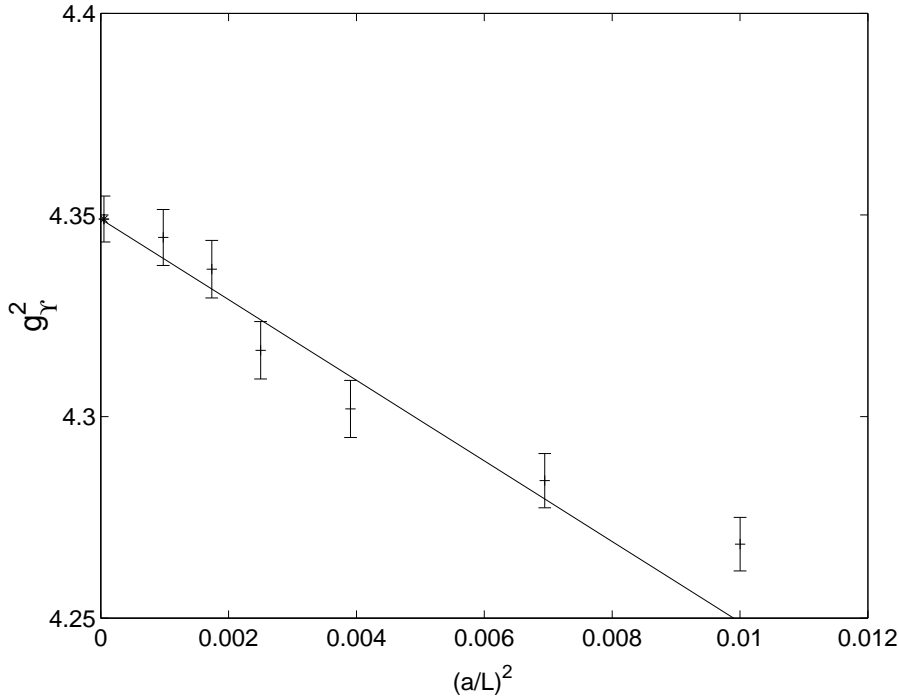}
  \end{center}
\caption{Continuum extrapolation of $\gup^2$ for $\gbar^2=2.4596$
as a typical example.}
\end{figure}
\begin{table}[!ht]
 \begin{center}
 \begin{tabular}{|cl|}\hline
  $\gbar^2$ & $\hphantom{10.0}\gup^2$  \\ \hline
   0.7390  &   \hphantom{1}0.8388(6) \\
   0.8166  &   \hphantom{1}0.9386(6) \\
   0.9176  &   \hphantom{1}1.0701(7) \\
   1.0595  &   \hphantom{1}1.2631(9) \\
   2.0373  &   \hphantom{1}3.0407(33)\\
   2.4596  &   \hphantom{1}4.349(6)  \\
   3.0280  &   \hphantom{1}7.210(15) \\
   3.7692  &              14.234(20) \\ \hline
 \end{tabular}
  \caption{Nonperturbative relation between $\gbar^2$ and $\gup^2$.}
  \label{contdata}
 \end{center}
\end{table}
\begin{figure}\label{coupling}
  \begin{center}
    \includegraphics[width=12cm]{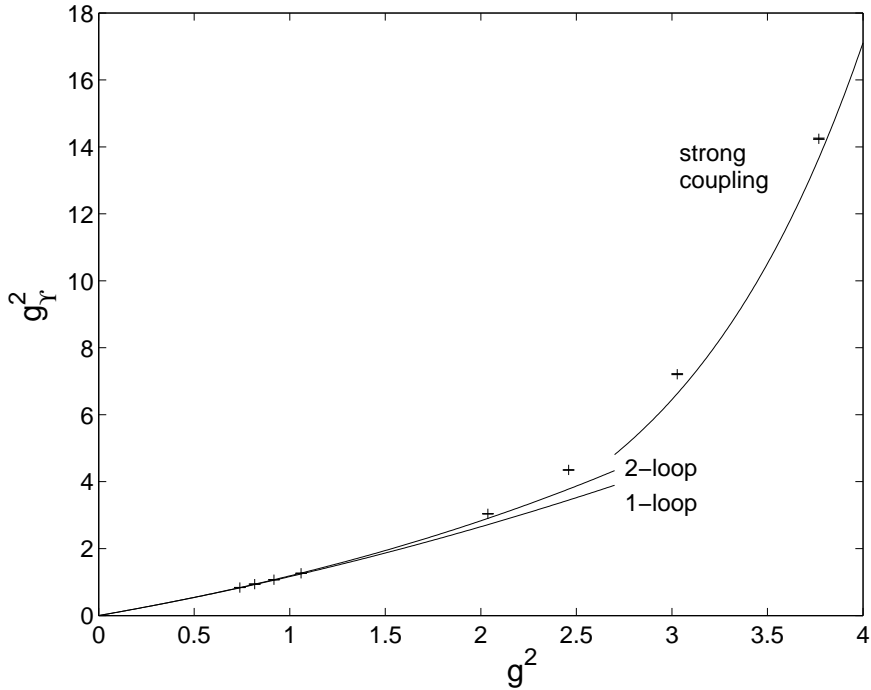}
  \end{center}
\caption{Coupling $\gup^2$ versus $\gbar^2$ with asymptotic expansions.}
\end{figure}
\begin{figure}\label{zoom}
  \begin{center}
    \includegraphics[width=12cm]{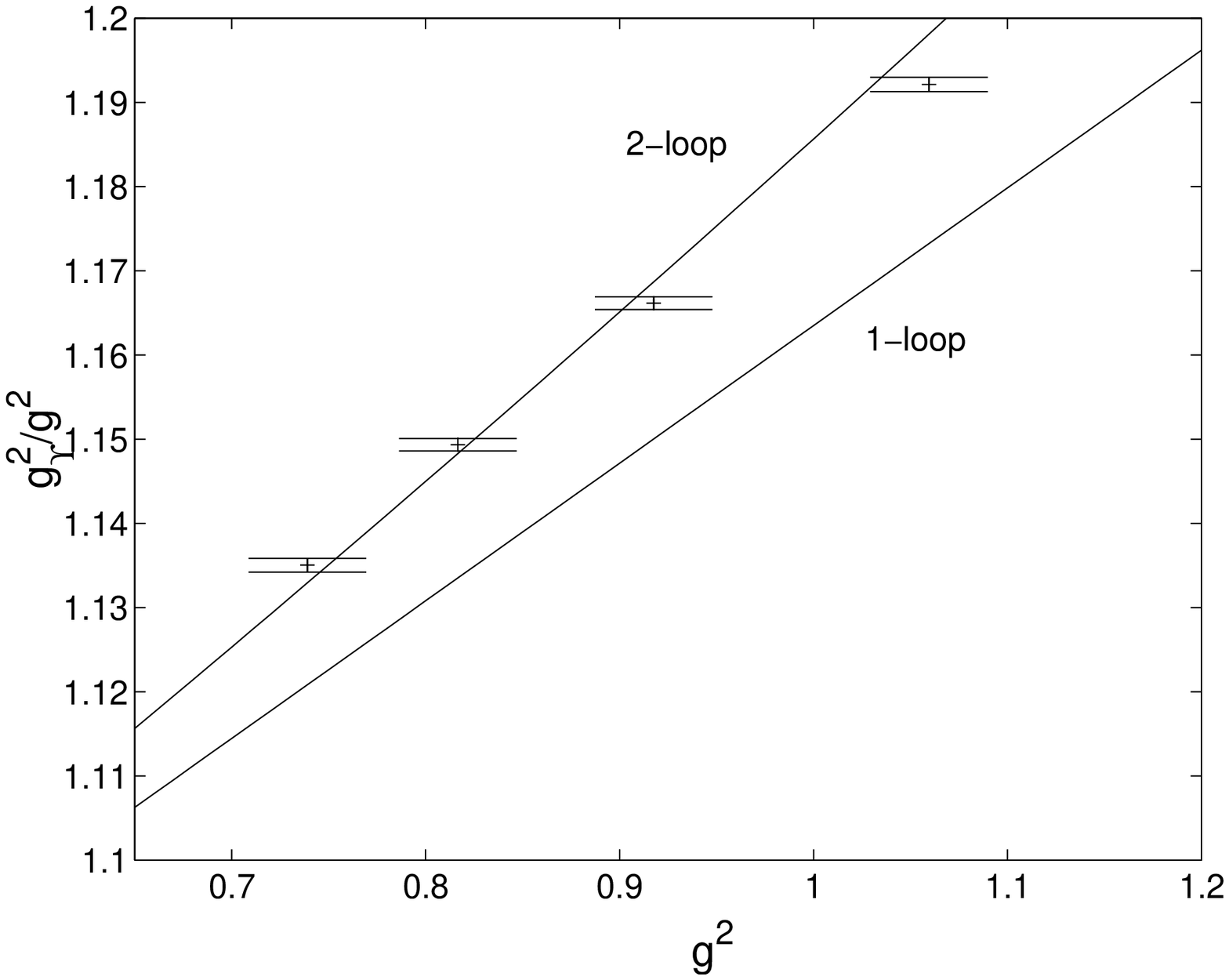}
  \end{center}
\caption{Close-up of $\gup^2$/$\gbar^2$ in the transition region.}
\end{figure}
A typical case is shown in
Fig.~2 where the coarsest lattice was not included in the fit.
Recently there has been some debate in the literature \cite{HN,HERESY} whether
the use of an asymptotic cutoff dependence proportional to $a^2$, based on
perturbation theory, is justified. Our data are compatible with this form, and
we {\em assume} it for continuum extrapolations in this work.
As a variation,
a linear cutoff dependence, as suggested for other quantities
in \cite{HN}, could however
also be fitted to the data in Fig.~2 and would lead to an exptrapolation
to $4.377(7)$, which differs not dramatically but clearly significantly.
To distinguish numerically between these and yet other behaviors
would require much higher precision and is of interest
for future study.
All our extrapolated continuum results are collected in Table~\ref{contdata}.
They are plotted in Fig.~3 and the close-up in Fig.~4, where one
sees the data branch off form the 2-loop curve (\ref{gugb}). 
After a very narrow
transition region they approach the strong coupling behavior 
which we constructed by performing the sum in (\ref{Upsc}).

To get a feeling for the lattice artefacts associated with our two
couplings individually
we estimated their step scaling function (SSF) for 
a pair of values corresponding to similar $L$,
$\gbar^2 \approx 1.06$
and $\gup^2 \approx 1.29$. The SSF --- the focus of interest in \cite{LWW} ---
gives $\gbar^2(2L)$ as a function of $\gbar^2(L)$.
Pairs of simulations at identical $\beta$ and sizes $L/a$ and $2L/a$ yield
\be
\Sigma(u,a/L)=\left. \gbar^2(2L) \right|_{\gbar^2(L)=u}.
\ee
$\Sigma$ is then extrapolated to the continuum limit.
A completely analogous quantity $\Sigma_\Upsilon$ is defined in terms
of $\gup^2$.
\begin{figure}\label{SSF}
  \begin{center}
    \includegraphics[width=12cm]{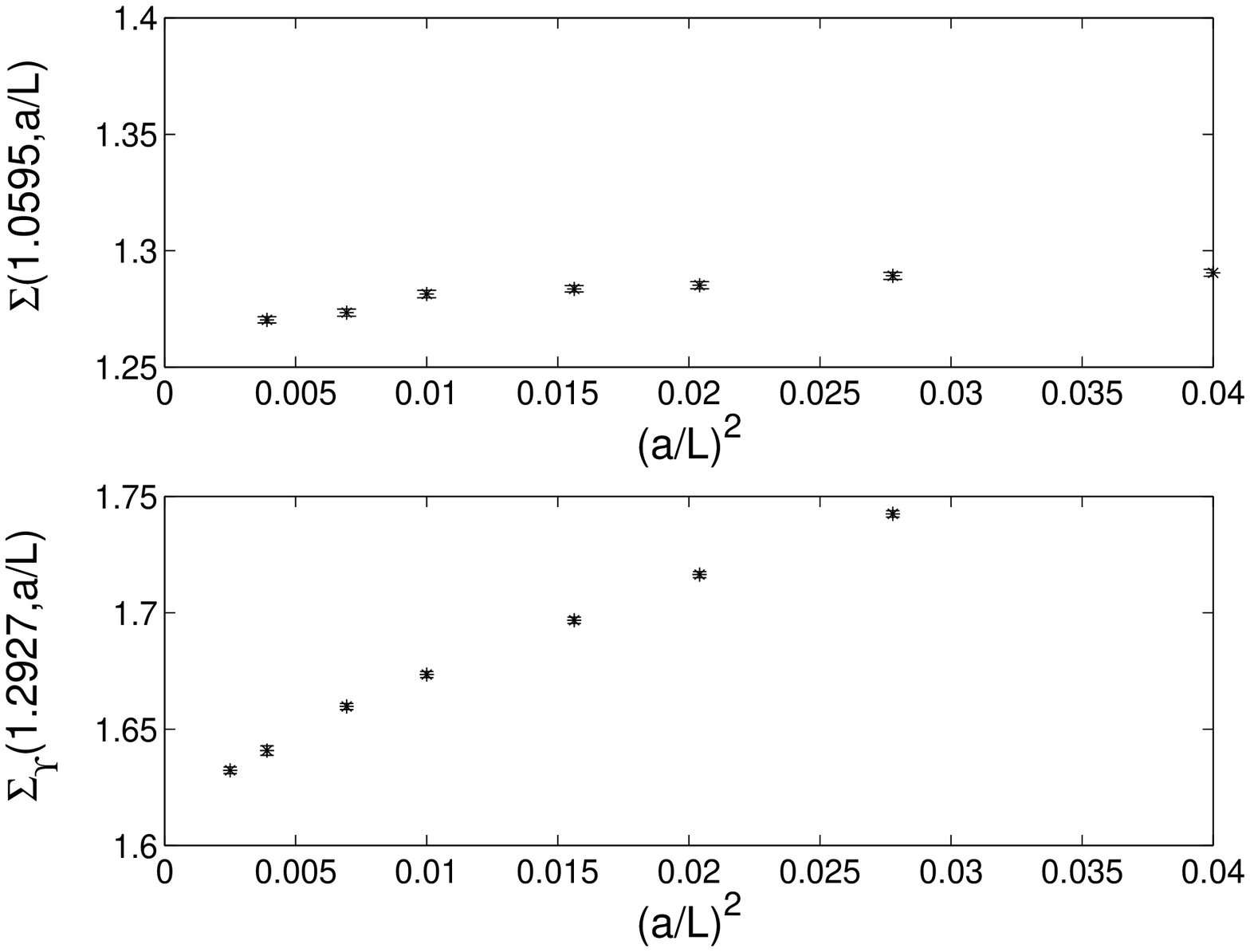}
  \end{center}
\caption{Continuum approach of step scaling functions.}
\end{figure}
In Fig.~5 the continuum approaches of $\Sigma$ and
$\Sigma_\Upsilon$ are shown. Lattice artefacts amount to a few
percent at say $L/a=8$ with $\Sigma_\Upsilon$ deviating
significantly more from its continuum limit than
$\Sigma$. This trend is expected, since $\Upsilon$
receives contributions form several low-lying eigenstates of the
transfer matrix (c.f. (\ref{Upstates})) while $\gbar$ is constructed
in terms of the massgap only.

Instead of relating the two couplings at one scale $L$ one may also
consider the connection between $\gup^2(L)$ and 
$\gbar^2(sL)$. 
This has already been incorporated in the perturbative formulas (\ref{gugb}).
One may hope that an
appropriate choice of $s$ improves the accuracy of the approximation
as was for instance found in \cite{SFSU2}.
A somewhat natural value is $s=2.7936 \, (n=3)$, for which $d_1(s)$ 
in (\ref{d1s}) vanishes and which coincides with the ratio of the
$\Lambda$-parameters associated with the short distance behaviors
of the two couplings. To compare such an expansion with non-perturbative
results we would have to simulate series of lattices of size $L$
and $sL$ at the same bare couplings and take the continuum limit.
While this is not possible, we gained good control over the
step scaling function for $\gbar^2(L)$ in \cite{LWW}.
We used its four loop approximation \cite{DSS,DSS1}\footnote{
We evaluated $\tilde{b}_4=0.0040$ at $n=3$ for eq.(3.47) in \cite{DSS1}.}
and fitted the remainder with an effective five loop coefficient to
evolve from $\gbar^2(L)$ to $\gbar^2(sL)$. In this way
we found however no
value $s\not= 1$ which significantly improved the series for $\gup^2(L)$.

To conclude, we have investigated two nonperturbatively
defined coupling constants for the O(3) nonlinear $\sigma$--model
with exponentially different low energy behaviors. Analytical
relations, valid  in the continuum limit,
 are available for both
weak and strong coupling. 
Precise numerical simulations covered
the intermediate range and matched with both asymptotic expansions.
As expected, the helicity modulus coupling shows larger lattice
artefacts than the finite volume massgap.

\noindent {\bf Acknowledgement} We would like to thank Erhard Seiler
for pointing out to us that the strong coupling asymptotics in an earlier
version of this paper was incorrect.

\begin{appendix}
\section{Two loop expansion of $\Upsilon$}

In this appendix we use lattice units with $a=1$ and take $T=L$.
For the perturbative expansion of the $\sigma$--model
on a finite lattice we have to fix the global O($n$)
invariance by the Fadeev-Popov technique \cite{PH}.
For O($n$) invariant integrands $I(s)$ it amounts
to the replacement $I(s) \to I(s) f(s) F(s)$ which does not
change the value of the integral. For the noninvariant function $f$
we take
\be
f(s)=\delta(M_1) \delta(M_2) \cdots \delta(M_{n-1})
\ee
where $M = \sum_x s(x)$ is the total magnetization,
which is hence forced to point in the $n$-direction.
The compensating Fadeev-Popov factor is in this case given by
\be
F(s) = |M|^{n-1}
\ee
up to an irrelevant overall constant factor.
The spins are parameterized by an $n-1$ component real field $\phi(x)$ 
\be
s=(g_0 \, \phi , \chi), \quad \chi=\sqrt{1-g_0^2 \phi^2}.
\ee
The resulting contributions are gathered in an effective action
\be
S_{\rm eff.} = \frac12 \sum_{x \mu} \left[
(\Delta_{\mu} \phi)^2 + \frac{1}{g_0^2} (\Delta_{\mu} \chi)^2
\right] + \sum_x \ln\chi - (n-1)\ln(\sum_x \chi).
\ee
It is understood to be expanded in $g_0$, 
\be
S_{\rm eff.} = \sum_{k=0}^\infty g_0^{2k} S^{(k)},
\ee
and the function $f$ still
has to be included in the path integral. It leads
to the omission of the zero momentum mode in 
the propagator of the $\phi$ field
and makes perturbative coefficients now well defined. The last two terms
in $S_{\rm eff.}$ correspond to the O($n$) invariant measure and to $F$.
We shall only need the leading terms
\bea
S^{(0)} &=& \frac12 \sum_x (\Delta_{\mu} \phi)^2 \label{S0}\\
S^{(1)} &=& \frac18 \sum_x (\Delta_{\mu} \phi^2)^2
-\frac12 \left(1-(n-1)\frac{1}{V}\right)\sum_x \phi^2.
\eea
The term $S^{(0)}$ defines the propagator
\be
\langle \phi_k(x) \phi_l(y) \rangle_0 = \delta_{kl} \, G(x-y) =
\delta_{kl} \, \frac{1}{V}\ {\sum_p}' \; \frac{\re^{ip(x-y)}}{\phat^2}
\ee
where $1\le k,l \le n-1$ and
the primed sum is over the appropriate lattice momenta excluding
$p=(0,0)$ and we have introduced $\phat_\mu = 2 \sin(p_\mu/2)$.

To compute $\Upsilon_1$ we set
\bea
s(x)s(x+a\hat{0}) &=& 1 - g_0^2 E^{(0)} - g_0^4 E^{(1)} + \rO(g_0^6)\\
E^{(0)} &=& \frac12 (\Delta_0 \phi)^2\\
E^{(1)} &=& \frac18 (\Delta_0 \phi^2)^2
\eea
and find
\be
\Upsilon_1 = \frac{2}{n g_0^2} \left(
1- g_0^2 \langle E^{(0)} \rangle_0 - g_0^4 \, \left[\langle E^{(1)} \rangle_0
-\langle E^{(0)} S^{(1)}  \rangle_0^c \, \right]
\right)
\ee
with the connected correlation in the last bracket.
These contributions evaluate to
\bea
\langle E^{(0)} \rangle_0 &=& \frac{n-1}{4} \left( 1-\frac{1}{L^2} \right)\\
\langle E^{(1)} \rangle_0 - \langle E^{(0)} S^{(1)}  \rangle_0^c
&=& \frac{n-1}{32}\left( 1-\frac{1}{L^2} \right)^2 -\frac{(n-1)(n-2)}{4}
A_1 \frac{1}{L^2}. \qquad
\eea

The contribution $\Upsilon_2$ has been expanded in (\ref{U2expand}).
We introduce
\be
\frac{1}{V g_0^4} 
\sum_{i<j}\sum_{x y}\, j^0_{ij}(x)j^0_{ij}(y) =
H^{(0)} + g_0^2 H^{(1)} + \rO(g_0^4)
\ee
\bea
H^{(0)} &=& \frac{4}{V} \sum_{k<l} \left(
\sum_x \phi_k \Delbar_0 \phi_l  \right)^2\\
H^{(1)} &=& \frac{1}{V} \sum_{k} \left(
\sum_x \phi^2 \Delbar_0 \phi_k  \right)^2
\eea
with the symmetric derivative
\be
\Delbar_\mu \phi(x) = \frac{1}{2} \left[\phi(x+\hat{\mu})
-\phi(x-\hat{\mu})\right]
\ee
and find
\be
\Upsilon_2 = \frac{2}{n(n-1)} \left( \langle H^{(0)} \rangle_0 + g_0^2 \, 
\left[\langle H^{(1)} \rangle_0
-\langle H^{(0)} S^{(1)}  \rangle_0^c \, \right] \right).
\ee
Numerical values are
\bea
\langle H^{(0)} \rangle_0 &=& (n-1)(n-2) (A_1-\frac14 A_2) \\
\langle H^{(1)} \rangle_0 &=& (n-1)[(n-2) B_1 + B_2] \\
\langle H^{(0)} S^{(1)}  \rangle_0^c &=& 
(n-1)(n-2) \Bigl[(A_1-\frac14 A_2) (A_1+\frac14 A_2 -\frac12(1-\frac{1}{V}))
\nonumber \\
&\phantom{=}& + \, 2 \, \frac{n-2}{V} A_3\Bigr]
\eea
In the above expressions the following $L$-dependent constants were
introduced,
\bea
A_1 &=& \frac{1}{V} {\sum_p}' \frac{1}{\phat^2}\\
A_2 &=& \frac{1}{V} {\sum_p}' \frac{\sum_\mu \phat_\mu^4}{(\phat^2)^2}\\
A_3 &=& \frac{1}{V} {\sum_p}' \frac{\sum_\mu \sin^2(p_\mu)}{(\phat^2)^3}\\
B_1 &=& -\sum_{x \mu} G(x)^2 \Delbar_\mu \Delbar_\mu G(x) \\
B_2 &=&  B_1 -2 \sum_{x \mu} G(x)  [\Delbar_\mu G(x)]^2 \, .
\eea
Evaluated as $x$- or $p$-sums as they stand, only $\rO(V)$ 
terms have to be summed.

It is now straightforward to express the coefficients
$k_1,k_2$ in
\be
\gup^2 =\frac{2}{n \Upsilon}  =g_0^2 + k_1 g_0^4 + k_2 g_0^6 + \rO(g_0^8)
\ee
in terms of the above constants. We evaluated them numerically
for $L=8 \ldots 100$ and determined the asymptotic behavior as
explained in the appendix of Ref.~\cite{BWW}.
The result is
\bea
k_1 &=& (n-2)\left[\frac{\ln(L)}{2\pi}
-0.12165689529\right] +  \frac{n-1}{4} +  \rO(1/L^2) \qquad\\
k_2-k_1^2 &=& (n-2)\left[\frac{\ln(L)}{(2\pi)^2}
+0.02514054821  \right] \nonumber \\[0.5ex]
&\phantom{=}& -(n-2)^2 \; 0.00773389318 + \frac{5}{96} +  \rO(1/L^2)
\eea
Errors are beyond the digits given here, and the coefficients
of $\ln(L)/L^2$ and $1/L^2$ 
corrections are also known but not listed here. They are of the same
size as the constants appearing here. To obtain the last fraction we
set $B_2=1/48 +\rO(1/L^2)$ which we observed to very high accuracy.

The massgap coupling (\ref{LWWcoupling}) has been computed
perturbatively up to two loops  in \cite{LWW} and to three loops
in \cite{DSS,DSS1}.
From the last reference we extract
\be
\gbar^2 = g_0^2 + m_1 g_0^4 + m_2 g_0^6 + \rO(g_0^8)
\ee
with
\bea
m_1 &=& (n-2)\left[\frac{\ln(L)}{2\pi} 
+\frac{\ln(\sqrt{2}/\pi)+\gamma}{2\pi}\right] +  \frac{1}{4} +  \rO(1/L^2) \qquad\\
m_2-m_1^2 &=& (n-2)\left[\frac{\ln(L)}{(2\pi)^2}
+ 0.021982285645  \right] + \frac{5}{96}  +  \rO(1/L^2).
\eea
The exact fraction was again found to numerical precision.

\end{appendix}

\end{document}